\documentclass[preprint,showpacs,preprintnumbers,amsmath,amssymb,nofootinbib]{revtex4}
\usepackage{graphicx,subfigure,textcomp}
\usepackage{dcolumn}
\usepackage{bm}
\usepackage{mathrsfs,slashed,amsmath}
\usepackage{natbib}
\usepackage{dsfont}
\allowdisplaybreaks
\begin{document}
\title{Finite Volume corrections to $\langle x\rangle_{u\pm d}$}
\author{Ludwig Greil}
\affiliation{Institut f\"ur Theoretische Physik, Universit\"at Regensburg, D-93040 Regensburg, Germany}
\author{Philipp Wein}
\affiliation{Institut f\"ur Theoretische Physik, Universit\"at Regensburg, D-93040 Regensburg, Germany}
\author{Peter C.~Bruns}
\affiliation{Institut f\"ur Theoretische Physik, Universit\"at Regensburg, D-93040 Regensburg, Germany}
\author{Andreas Sch\"afer}
\affiliation{Institut f\"ur Theoretische Physik, Universit\"at Regensburg, D-93040 Regensburg, Germany}
\date{\today}
\begin{abstract}
In this paper we calculate the full one-loop finite volume corrections to the quantities $\langle x \rangle_{u\pm d}$ within the framework of two-flavor baryon chiral perturbation theory. For the isovector case, we show estimates of these effects for the leading one-loop corrections based on fits carried out in previously published works.
\end{abstract}
\maketitle
\section{Introduction}
One of the major challenges in nuclear physics today is to understand the structure of the nucleon arising from QCD dynamics. Major progress is made by the continuous improvement of lattice QCD, see \cite{Durr:2008zz,Hagler:2009ni,Fodor:2012gf,Durr:2011mp}. However, in some cases results obtained from lattice simulations do not seem to extrapolate naturally to the experimentally known results. The most serious such discrepancy is observed for the first moments of the isovector quark distribution function of the nucleon, e.g. $\langle x\rangle_{u-d}$, for which high precision data is available \cite{Bali:2012av,Bali:2014}.\\The computation of observables on the lattice in general suffers from a number of systematic uncertainties: Both lattice spacing and lattice volume are finite, and up to recently most simulations used quark masses that are much larger than the physical ones. Thus the quality of lattice results depends on the control over a threefold extrapolation: the continuum extrapolation ($a\rightarrow0$), the extrapolation to the thermodynamic limit ($V\rightarrow\infty$) and the chiral extrapolation ($m_q^{\text{latt}}\rightarrow m_q^{\text{phys}}$). While the last two extrapolations nowadays are becoming more and more obsolete due to almost physical quark masses and large volumes that are used in lattice simulations, the continuum extrapolation is still problematic, also because for lattice constants below $0.05\,\text{fm}$ one encounters very long topological autocorrelation times. The latter problem can be avoided by using open boundary conditions \cite{Luscher:2011kk,Luscher:2012av}.\\Possible finite volume corrections can be treated within the framework of chiral perturbation theory (ChPT) \cite{Gasser:1986vb,Gasser:1987ah,Gasser:1987zq,Hasenfratz:1989pk}. Most of the time when finite volume effects have been analyzed, they have been found to be around $5-10\%$ \cite{Ali Khan:2003cu,Greil:2011aa} and thus too small to explain the observed discrepancies. Nonetheless, a thorough analysis of lattice QCD data should incorporate these corrections. This work constitutes a covariant calculation of the finite volume effects, for the heavy baryon result, see ref.~\cite{Detmold:2005pt}.\\In this paper, we present the finite volume corrections to the quantities $\langle x\rangle_{u\pm d}$ within the framework of $SU(2)_f$ covariant baryon chiral perturbation theory (BChPT) to full one loop order. The paper is structured as follows: in section~\ref{sec:setup} we give a short overview of the ChPT setup for this particular case, in sec.~\ref{sec:calc} we present our finite volume calculation and in sec.~\ref{sec:est} we give a rough estimate of the size of these finite volume corrections. We give a short conclusion in sec.~\ref{sec:conc}.
\section{ChPT setup\label{sec:setup}}
\begin{figure}
\centering
\subfigure[]{\includegraphics[width=0.17\textwidth]{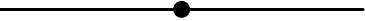}}\hspace{0.5cm}\subfigure[]{\includegraphics[width=0.17\textwidth]{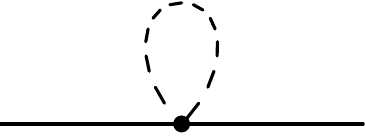}}\hspace{0.5cm}\subfigure[]{\includegraphics[width=0.17\textwidth]{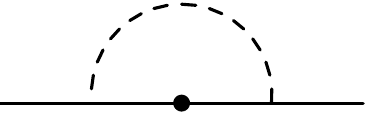}}\\\subfigure[]{\includegraphics[width=0.17\textwidth]{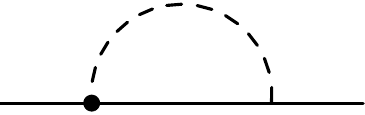}}\hspace{0.5cm}\subfigure[]{\includegraphics[width=0.17\textwidth]{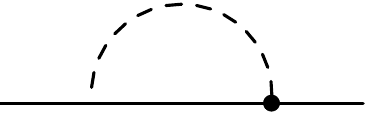}}\hspace{0.5cm}\subfigure[]{\includegraphics[width=0.17\textwidth]{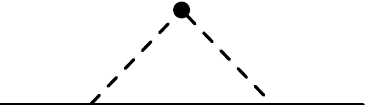}}
\caption{We show all graphs contributing to $A_{2,0}^{u\pm d}(t=0)$ at full one-loop level, where the solid line denotes a nucleon, the dashed line represents a pion and the dots represent an operator insertion.}
\label{fig:graphs}
\end{figure}This paper mostly concerns itself with calculating the finite volume corrections to $\langle x \rangle_{u\pm d}$ and thus, for general introductions to BChPT we refer the reader to the reviews \cite{Scherer:2012xha,Kubis:2007iy}. The full one-loop calculation for the nucleon GPDs in $SU(2)_f$ has been performed in detail in \cite{Wein:2014wma}. Here, we will only cite the results needed for the calculation of the finite volume corrections to $\langle x \rangle_{u-d}$, all other details can be found in said reference.\\We are interested in analyzing the matrix elements of twist-two quark operators including one covariant derivative, i.e.
\begin{align}
\langle p'\left|\mathcal{O}^q_{\mu\nu}\right|p\rangle,\qquad \mathcal{O}^q_{\mu\nu}=\frac{i}{2}\bar{q}\gamma_{\{\mu}\overleftrightarrow{D}_{\nu\}}q,\qquad \overleftrightarrow{D}_{\mu}=\overrightarrow{D}_{\mu}-\overleftarrow{D}_{\mu}.\label{eq:operators}
\end{align}
Here, $p'$ and $p$ are the outgoing (incoming) nucleon four-momenta and $_{\{\ldots\}}$ denotes total symmetrization of indices and subtraction of traces. These matrix elements are connected to so called generalized form factors $A^q_{2,0}(t)$, $B^q_{2,0}(t)$ and $C^q_{2,0}(t)$ in the following manner:
\begin{align}
\langle p'\left|\mathcal{O}^q_{\mu\nu}\right|p\rangle=\bar{u}(p')\left[\gamma_{\{\mu}\bar{p}_{\nu\}}A^q_{2,0}(t)-\Delta^{\alpha}\sigma_{\alpha\{\mu}\bar{p}_{\nu\}}\frac{iB_{2,0}^q(t)}{2m_N}+\Delta_{\{\mu}\Delta_{\nu\}}\frac{C_{2,0}^q(t)}{m_N}\right]u(p).
\end{align}
Here, we have introduced two kinematical variables, namely $\bar{p}=(p'+p)/2$, $\Delta=p'-p$ and $t=\Delta^2$. The generalized form factors $A^q_{2,0}(t)$, $B^q_{2,0}(t)$ and $C^q_{2,0}(t)$ are connected to the GPDs $H^q(x,\xi,t)$ and $E^q(x,\xi,t)$, which are defined and discussed in \cite{Ji:1996ek,Diehl:2003ny}. In the forward limit, the form factor $A^q_{2,0}(0)$ is connected to the first moments of the parton distribution functions (PDFs) $q(x)$ and $\bar{q}(x)$:
\begin{align}
\langle x\rangle_q=\int_{0}^1dx\,x\left[q(x)+\bar{q}(x)\right]=A_{2,0}^q(0).
\end{align}
Since we want to access the quantity $\langle x\rangle_q$ via $SU(2)_f$ covariant BChPT, we employ the decomposition of the operators in eq.~\eqref{eq:operators} into chiral fields presented in \cite{Wein:2014wma}. This decomposition enables us to calculate the infinite volume value for $A_{2,0}^q(t)$ to full one-loop order and then extract the infinite volume value for the first moment of the PDAs. This has been done in \cite{Wein:2014wma} using the infrared regularization scheme \cite{Becher:1999he}. We show the contributing one-loop graphs in fig.~\ref{fig:graphs}. We present the infinite volume result for each diagram seperately for $A_{2,0}^{u+ d}(t=0)$ in eqs.~\eqref{eq:infvolressbeg}-\eqref{eq:infvolressend} and the results for $A_{2,0}^{u-d}(t=0)$ in eqs.~\eqref{eq:infvolresbeg}-\eqref{eq:infvolresend}:
\begin{align}
A_{2,0}^{u+d,a}(t=0)&=8 \mathcal{Z} l^s_{0,1} + 32 M_{\pi}^2 l^s_{2,2},\label{eq:infvolressbeg}\\
A_{2,0}^{u+d,b}(t=0)&=0,\\
\begin{split}
A_{2,0}^{u+d,c}(t=0)&=\frac{6 g_A^2 l^s_{0,1}}{F_{\pi}^2} \Bigl(-4 (d-2) m_N^4 (I_{12}^{(2)}-I_{12}^{(4)})+I_{10}^{(0)}-4 m_N^2 (I_{11}^{(1)}-I_{11}^{(3)})\\
               &\quad+4 m_N^4 (I_{12}^{(3)}-I_{12}^{(5)})\Bigr), 
\end{split}\\
\begin{split}
A_{2,0}^{u+d,d+e}(t=0)&=\frac{24 g_A}{F_{\pi}^2} \biggl(2 m_N^2 \Bigl(2 m_N l^s_{1,18} \bigl(2 (d-1) I_{11}^{(2)}-d I_{11}^{(4)}\bigr)\\
                      &\quad+l^s_{1,15} \bigl(2 I_{10}^{(2)}+M_{\pi}^2 (4 I_{11}^{(0)}-4I_{11}^{(1)}+I_{11}^{(3)})\bigr)\Bigr)\\
                      &\quad+l^s_{1,13} \Bigl(2 M_{\pi}^2 \bigl(I_{10}^{(0)}+m_N^2 (I_{11}^{(3)}-2 I_{11}^{(1)})\bigr)+2 I_{10}^{(2)} m_N^2\\
                      &\quad+4 m_N^4 (2(I_{11}^{(2)}+I_{11}^{(3)}-I_{11}^{(4)})-I_{11}^{(5)})\Bigr)-2 I_{11}^{(3)} m_N^2 l^s_{1,6}\biggr), 
\end{split}\\
A_{2,0}^{u+d,f}(t=0)&=-\frac{48 g_A^2 l^s m_N^4 (2 I_{21}^{(4)}+I_{21}^{(5)})}{F_{\pi}^4},\label{eq:infvolressend}\\
A_{2,0}^{u-d,a}(t=0)&=4l_{0,1}\mathcal{Z}+16l_{2,2}M_{\pi}^2,\label{eq:infvolresbeg}\\
A_{2,0}^{u-d,b}(t=0)&=-\frac{4l_{0,1}}{F_{\pi}^2}I_{10}^{(0)},\\
\begin{split}
A_{2,0}^{u-d,c}(t=0)&=-\frac{g_A^2 l_{0,1}}{F_{\pi}^2} \Bigl(4 (2-d) m_N^4 (I^{(2)}_{12}-I^{(4)}_{12})+I_{10}^{(0)}-4 m_N^2 (I_{11}^{(1)}-I_{11}^{(3)})\\
&\quad+4 m_N^4 (I_{12}^{(3)}-I_{12}^{(5)})\Bigr),
\end{split}\\
\begin{split}
A_{2,0}^{u-d,d+e}(t=0)&=\frac{g_A}{F_{\pi}^2} \biggl(4 l_{0,2} \Bigl(I_{10}^{(0)}-2 I_{11}^{(1)}m_N^2+I_{11}^{(3)}m_N^2\Bigr)-8I_{11}^{(3)}m_N^2 (l_{1,6}+l_{1,7})\\
&\quad+4m_N^3 \left(l_{1,1}-2 l_{1,3}\right) (-2I_{11}^{(2)}-2I_{11}^{(3)}+2 I_{11}^{(4)}+I_{11}^{(5)})\\
&\quad+16 m_N^3 (l_{1,18}+l_{1,19}) (2 (d-1) I_{11}^{(2)}-d I_{11}^{(4)})\\
&\quad+4 m_N l_{1,8} \Bigl(m_N^2 \bigl(-8 I_{10}^{(2)}+2 M_{\pi}^2 (4 I_{11}^{(1)}-4I_{11}^{(2)}-4I_{11}^{(3)}+2I_{11}^{(4)}+I_{11}^{(5)})\\
&\quad-4 m_N^2 (4I_{11}^{(2)}+4I_{11}^{(3)}-12I_{11}^{(4)}-4I_{11}^{(5)}+2I_{11}^{(6)}+5I_{11}^{(7)}+I_{11}^{(8)})\bigr)\\
&\quad-4 I_{10}^{(0)}M_{\pi}^2\Bigr)+(l_{1,13}+l_{1,14})\Bigl(8 I_{10}^{(0)}M_{\pi}^2+8 m_N^2 \bigl(I_{10}^{(2)}+M_{\pi}^2 (I_{11}^{(3)}-2I_{11}^{(1)})\\
&\quad+2 m_N^2 (2(I_{11}^{(2)}+I_{11}^{(3)}-I_{11}^{(4)})-I_{11}^{(5)})\bigr)\Bigr)\\
&\quad+8 m_N^2 (l_{1,15}+l_{1,16})\left(2I_{10}^{(2)}+M_{\pi}^2 (4 I_{11}^{(0)}-4I_{11}^{(1)}+I_{11}^{(3)})\right)\biggr).
\end{split}\\
A_{2,0}^{u-d,f}(t=0)&=0.\label{eq:infvolresend}
\end{align}
All infrared integrals $I^{(i)}_{nm}(p^2,m^2,M^2)$ that appear in eqs.~\eqref{eq:infvolressbeg}-\eqref{eq:infvolresend} are to be evaluated at the point $(p^2,m^2,M^2)=(m_N^2,m_N^2,M_{\pi}^2)$. Their definitions are given in apps.~\ref{app:integrals} and \ref{app:tensor}. In eq.~\eqref{eq:infvolresbeg} we have used the quantity $\mathcal{Z}$, which stands for the nucleon wave function renormalization constant and which in terms of infrared integrals takes the form
\begin{align}
\mathcal{Z}=1+\frac{3g_A^2}{4F_{\pi}^2}\frac{\partial}{\partial \slashed{p}}\Bigl[M_{\pi}^2(\slashed{p}+m_N)I^{(0)}_{11}+(m_N^2-\slashed{p}^2)\slashed{p}^2I^{(1)}_{11}-(m_N+\slashed{p})I^{(0)}_{10}\Bigr]\Bigr|_{\slashed{p}=m_N}
\end{align}
In accordance with the strategy of computing the finite volume effects in \cite{Wein:2011ix} we now have all the building blocks to carry out a similar calculation for $\langle x \rangle_{u\pm d}$.
\section{Finite volume corrections\label{sec:calc}}
When calculating the finite volume effects we rely on the seminal work published in \cite{Gasser:1986vb,Gasser:1987ah,Gasser:1987zq,Hasenfratz:1989pk}. This effective field theory formalism is based on the observation that the finite volume effects are predominantly caused by pions travelling around the box of finite extent. Thus, we rewrite our infinite volume quantity $\langle x\rangle_{u-d}^{\infty}$ as
\begin{align}
\langle x\rangle^{\infty}_{u-d}=\langle x\rangle_{u-d}^{\infty}-\langle x\rangle_{u-d}(L)+\langle x\rangle_{u-d}(L)\equiv\delta\langle x\rangle_{u-d}(L)+\langle x\rangle_{u-d}(L).
\end{align}
According to \cite{Hasenfratz:1989pk}, the finite volume corrections $\delta\langle x\rangle_{u-d}(L)$ are obtained by calculating the difference between integral and discrete sum over all possible loop momenta, i.e.
\begin{align}
\delta I_{nm}(L)=-i\int\frac{dq_0}{2\pi}\left[\int\frac{d^3\mathbf{q}}{(2\pi)^3}-\frac{1}{L_1L_2L_3}\sum_{\mathbf{q}}\right]_{\text{IR}}\frac{1}{(M_{\pi}^2-q^2)^m(m_N^2-(p-q)^2)^n}.
\end{align}
The subscript $_{\text{IR}}$ corresponds to extending the integration for Feynman parameters that combine meson- with nucleon-propagators from $[0,1]$ to $[0,\infty]$. Here, $L_i$ denotes the box-length in the $i$-th spatial direction and the sum is carried out for three-momenta $\mathbf{q}=2\pi(\frac{n_1}{L_1},\frac{n_2}{L_2},\frac{n_3}{L_3})^T$ for integer $n_i$. If one follows the steps described in \cite{Hasenfratz:1989pk} one finds for the relevant integrals
\begin{align}
\delta I_{10}(L)&=-\frac{1}{(2\pi)^2}\sum_{\mathbf{n}\neq0}\left(\frac{M_{\pi}^2}{\mathbf{L}_{\mathbf{n}}^2}\right)^{\frac{1}{2}}K_1\left(\sqrt{\mathbf{L}_{\mathbf{n}}^2M_{\pi}^2}\right),\\
\delta I_{01}(L)&=0,\\
\delta I_{11}(L)&=-\frac{1}{2(2\pi)^2}\int_0^{\infty}du\sum_{\mathbf{n}\neq0}K_0\left(\sqrt{\mathbf{L}_{\mathbf{n}}^2f(u,m_N^2)}\right)\exp\left\{iu\mathbf{p}\cdot\mathbf{L}_{\mathbf{n}}\right\}.
\end{align}
Here, we have introduced a quantity which is defined as $\mathbf{L}_{\mathbf{n}}=(n_1L_1,n_2L_2,n_3L_3)$. Furthermore we denote the $i$-th modified Bessel function of the second kind by $K_i$ and we have defined the function
\begin{align}
f(u,p^2)=um_N^2+(1-u)M_{\pi}^2+(u^2-u)p^2,
\end{align}
which naturally appears when one combines the nucleon- and meson-propagator. To obtain the finite volume corrections, we start from eqs.~\eqref{eq:infvolressbeg}-\eqref{eq:infvolresend} and we first reduce all infrared integrals appearing in these equations to the standard scalar integrals $I_{10}$, $I_{01}$ and $I_{11}$ using the relations from App.~\ref{app:tensor}. See however that there is in principle a problem with using tensor decomposition when operating in a finite volume due to breaking of Lorentz invariance \cite{Greil:2011aa}. Going from infinite volume to finite volume corrections corresponds to the replacement
\begin{align}
I_{10}\longrightarrow\delta I_{10},\qquad I_{10}\longrightarrow\delta I_{01},\qquad I_{10}\longrightarrow\delta I_{11}.
\end{align}
When we carry out this replacement, we find that the term proportional to $l_{2,2}$ drops out since this term does not recieve corrections due to pions travelling around the box. For the case of four space-time dimensions $d=4$ and isotropic lattices $L_1=L_2=L_3$, we end up with 
\begin{align}
\begin{split}
\delta\langle x\rangle_{u+d}(L)&=\int_{0}^{\infty}du\biggl[\alpha_{u+d}\left(K_0\left(L M_{\pi} \sqrt{\mathbf{n}^2}\right)+K_2\left(L M_{\pi}\sqrt{\mathbf{n}^2}\right)\right)\\
&\quad+\beta_{u+d} K_0\left(L \sqrt{\mathbf{n}^2f(u,m_N^2)}\right) \cos(Lu\mathbf{n}\cdot\mathbf{p})\\
&\quad+\gamma_{u+d} K_1\left(L \sqrt{\mathbf{n}^2 f(u,m_N^2)}\right) \cos (L u\mathbf{n}\cdot\mathbf{p})+\delta_{u+d}K_1\left(L M_{\pi} \sqrt{\mathbf{n}^2}\right)\biggr],
\end{split}\\
\begin{split}
\delta\langle x\rangle_{u-d}(L)&=\int_{0}^{\infty}du\biggl[\alpha_{u-d} K_1\left(L M_{\pi} \sqrt{\mathbf{n}^2}\right)+\beta_{u-d} K_0\left(L \sqrt{\mathbf{n}^2f(u,m_N^2)}\right) \cos(Lu\mathbf{n}\cdot\mathbf{p})\\
&\quad+\gamma_{u-d} K_1\left(L \sqrt{\mathbf{n}^2 f(u,m_N^2)}\right) \cos (L u\mathbf{n}\cdot\mathbf{p})\biggr],
\end{split}
\end{align}
where the single coefficients $\alpha_{u\pm d}$, $\beta_{u\pm d}$, $\gamma_{u\pm d}$ and $\delta_{u+d}$ take the following form:
\begin{align}
\alpha_{u+d}&=\frac{g_A^2 l^s M_{\pi}^4 \theta (1-u)\theta(u)}{2 \pi ^2 F_{\pi}^4 m_N^2},\\
\begin{split}
\beta_{u+d}&=-\frac{g_A M_{\pi}^2}{\pi ^2 F_{\pi}^4 m_N^3} \biggl(F_{\pi}^2 \Bigl(g_A m_N l^s_{0,1} \left(3 m_N^2-2 M_{\pi}^2\right)\\
           &\quad+2 \left(12 m_N^4-7 m_N^2 M_{\pi}^2+M_{\pi}^4\right)   \left(m_N l^s_{1,15}+l^s_{1,18}\right)\\
           &\quad+2 m_N l^s_{1,6} (m_N-M_{\pi}) (m_N+M_{\pi})\Bigr)+g_A l^s m_N \left(3 M_{\pi}^2-2m_N^2\right)\biggr)
\end{split}\\
\begin{split}
\gamma_{u+d}&=\frac{g_A^2 M_{\pi}^2 \sqrt{L^2 \mathbf{n}^2}}{2 \pi ^2 F_{\pi}^4 m_N^2 \sqrt{f(u,m_N^2)}} \biggl(F_{\pi}^2 u l^s_{0,1} \left(3 m_N^4 u-4 m_N^2 M_{\pi}^2+M_{\pi}^4\right)\\
            &\quad+l^sM_{\pi}^2 (u-1) (m_N^2-M_{\pi}^2)\biggr)
\end{split}\\
\begin{split}
\delta_{u+d}&=\frac{g_A M_{\pi}^3 \theta (1-u)\theta(u)}{\pi ^2 F_{\pi}^4 m_N^3 \sqrt{L^2 \mathbf{n}^2}} \biggl(F_{\pi}^2 \Bigl(m_N \bigl(2 g_A l^s_{0,1}+2 l^s_{1,15}  \left(9 m_N^2-2 M_{\pi}^2\right)\\
            &\quad-15 m_N^2 l^s_{1,13}+4l^s_{1,6}\bigr)+4 l^s_{1,18} \left(6 m_N^2-M_{\pi}^2\right)\Bigr)-5 g_A l^s m_N\biggr)
\end{split}\\
\begin{split}
\alpha_{u-d}&=\frac{M_{\pi}\theta (1-u)\theta(u)}{30 \pi^2 F_{\pi}^2 L m_N^3\sqrt{\mathbf{n}^2}} \biggl(10 l_{0,1} \left(3 \left(g_A^2+1\right) m_N^3-g_A^2 m_N M_{\pi}^2\right)\\
&\quad+g_A M_{\pi}^2 \Bigl(120 m_N^4 l_{1,8}-75 m_N^3 (l_{1,13}+l_{1,14})+90 m_N^3 (l_{1,15}+l_{1,16})-22 m_N^2 M_{\pi}^2 l_{1,8}\\
&\quad+5 l_{1,1} \left(3 m_N^2-M_{\pi}^2\right)+20 (l_{1,18}+l_{1,19}) \left(6m_N^2-M_{\pi}^2\right)-30 m_N^2 l_{1,3}\\
&\quad-20 m_N M_{\pi}^2 (l_{1,15}+l_{1,16})-10 m_N l_{0,2}+20 m_N l_{1,67}+6 M_{\pi}^4 l_{1,8}+10 M_{\pi}^2 l_{1,3}\Bigr)\biggr)
\end{split}\\
\begin{split}
\beta_{u-d}&=\frac{g_A M_{\pi}^2}{60 \pi ^2 F_{\pi}^2 m_N^3} \Bigl(10 g_A m_N l_{0,1} \left(3 m_N^2-2 M_{\pi}^2\right)-240 m_N^5 (l_{1,15}+l_{1,16})\\
&\quad+16 m_N^4 M_{\pi}^2 l_{1,8}+140 m_N^3 M_{\pi}^2 (l_{1,15}+l_{1,16})+10 l_{0,2} \left(4 m_N^3-m_N M_{\pi}^2\right)\\
&\quad-20 m_N^3 (l_{1,6}+l_{1,7})-28 m_N^2 M_{\pi}^4 l_{1,8}+20 m_N^2 M_{\pi}^2 l_{1,1}-40 m_N^2 M_{\pi}^2 l_{1,3}\\
&\quad-20 (l_{1,18}+l_{1,19}) \left(12 m_N^4-7 m_N^2 M_{\pi}^2+M_{\pi}^4\right)-20 m_N M_{\pi}^4 (l_{1,15}+l_{1,16})\\
&\quad+20 m_N M_{\pi}^2 l_{1,67}+6 M_{\pi}^6 l_{1,8}-5 M_{\pi}^4 l_{1,1}+10 M_{\pi}^4 l_{1,3}\Bigr) 
\end{split}\\
\gamma_{u-d}&=-\frac{g_A^2 L M_{\pi}^2 u l_{0,1} \left(m_N^4 (12-9 u)-4 m_N^2 M_{\pi}^2+M_{\pi}^4\right) \mathbf{n}^2}{12 \pi ^2 F_{\pi}^2 m_N^2   \sqrt{\mathbf{n}^2f(u,m_N^2)}}
\end{align}
Above we have introduced the quantity $\mathbf{n}=(n_1,n_2,n_3)^T$ in order to arrive at a short notation for the finite volume corrections. Technically speaking to arrive at the full expressions, one has to use that to the order we are working at, the nucleon mass $m_N$ itself has an expansion in $M_{\pi}^2$ \cite{Gasser:1987rb,Becher:1999he}, i.e.
\begin{align}
m_N=m_0-4c_1M_{\pi}^2+\mathcal{O}(p^3),
\end{align}
and hence, when implementing the finite volume corrections, $m_N$ has to be replaced with the above expression. Note that every time the mass quantity $M_{\pi}$ appears, it refers to the pion mass in the infinite volume limit, so $M_{\pi}\equiv M_{\pi}^{\infty}$.
\section{Estimate of finite volume effects\label{sec:est}}
\begin{figure}[tb]
\centering
\subfigure[]{\includegraphics[width=0.7\textwidth]{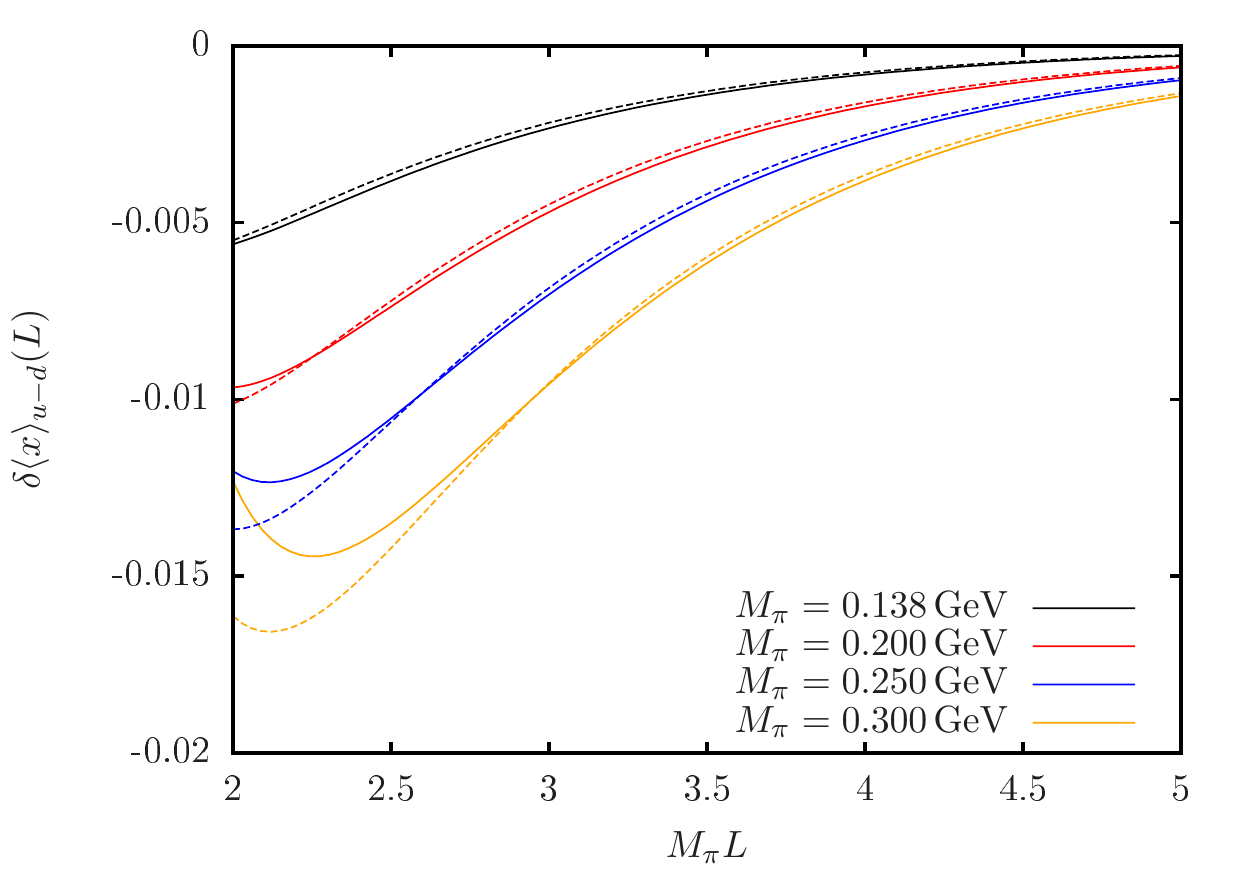}\label{fig:estimate_1}}\\\subfigure[]{\includegraphics[width=0.7\textwidth]{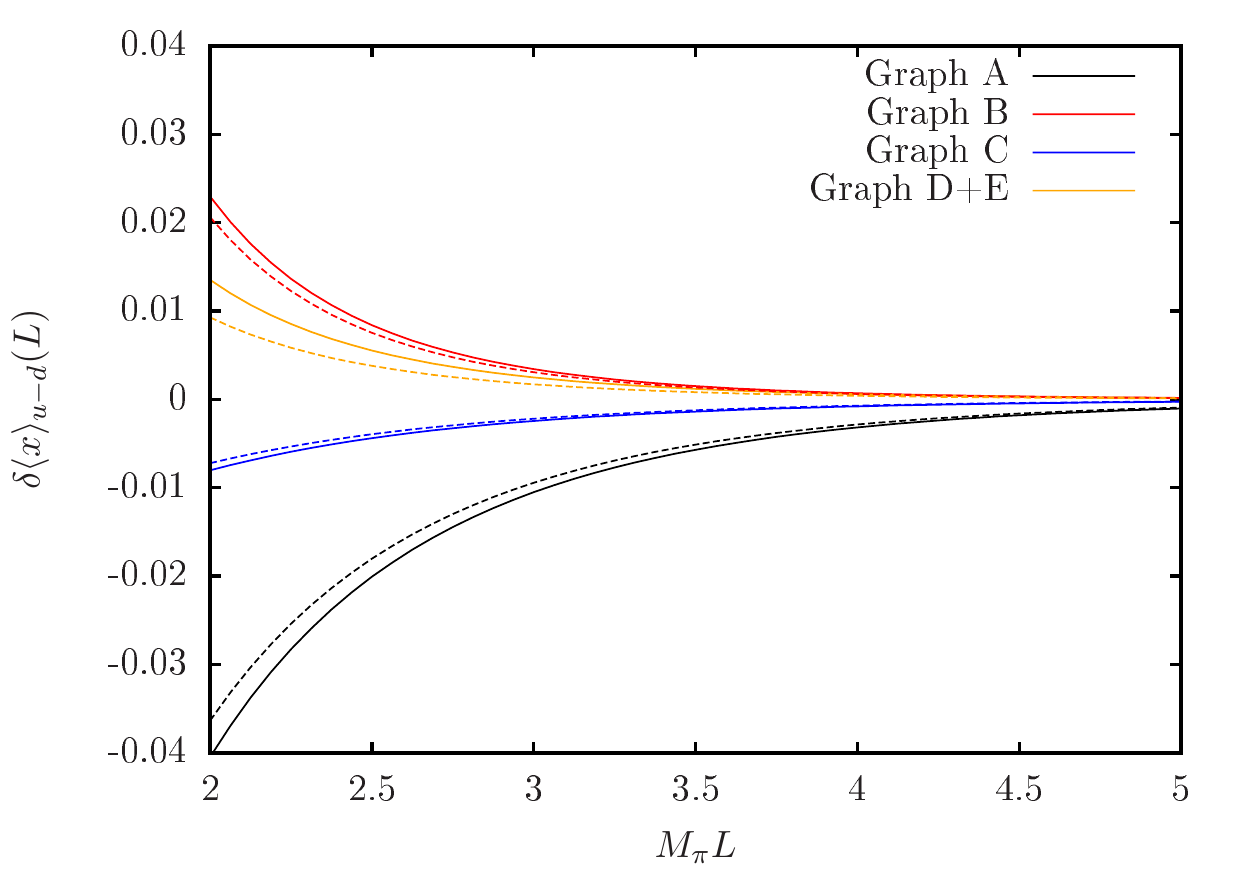}\label{fig:estimate_2}}
\caption{The plots above show estimates for the leading loop finite volume corrections to $\langle x\rangle_{u-d}$ for the input parameters defined in tab.~\ref{tab:fitparams}. In fig.~\ref{fig:estimate_2} we have set $M_{\pi}=250\,\text{MeV}$.}
\end{figure}

In this section, we want to investigate the leading one-loop finite volume corrections to the isovector case, because both the lattice data and the input parameters for the isosinglet case are very poorly determined, due to the importance of disconnected contributions. For the estimate of the magnitude we have to choose input parameters for the low energy constants that contribute at leading one-loop order, which are $l_{0,1}$, $l_{0,2}$ and $l_{2,2}$. We opt to use the fit results published in \cite{Dorati:2007bk}. When we compare the LECs defined in \cite{Wein:2014wma} with the definitions used in \cite{Dorati:2007bk} we find that
\begin{align}
l_{0,1}\equiv\frac{a_{2,0}^v}{4},\qquad l_{0,2}\equiv\frac{\Delta a_{2,0}^v}{4},\qquad l_{2,2}=\frac{c_8}{4m_0^2}.
\end{align}
Dorati et al. quote the values presented in table~\ref{tab:fitparams} as their fit results.
\begin{table}[tb]
\caption{Fit results taken from \cite{Dorati:2007bk} where $\Delta a_{2,0}^v$ was taken as an input value for the first set (full) and it was fitted in the second set (dashed).}
\begin{ruledtabular}
\begin{tabular}{c c c c}
set    & $a_{2,0}^v$      & $\Delta a_{2,0}^v$ & $c_8^r(1\,\text{GeV})$\\
\hline
full   & $0.157\pm 0.006$ & $0.21$ (fixed)     & $-0.283\pm0.011$\\
dashed & $0.141\pm0.0057$  & $0.144\pm0.034$    & $-0.213\pm0.03$
\end{tabular}
\label{tab:fitparams}
\end{ruledtabular}
\end{table}
We arrive at the leading one-loop corrections by setting most of the low energy constants to zero, in our case all LECs but $l_{0,1}$, $l_{0,2}$. The LEC $l_{2,2}$ does not contribute to the finite volume corrections since to this order it only appears at tree level. This leads us to the leading one loop formulae, which take the reduced form
\begin{align}
\begin{split}
\alpha_{u-d}^{\text{red}}&=\frac{M_{\pi}\theta (1-u)\theta(u)}{30 \pi^2 F_{\pi}^2 L m_N^3\sqrt{\mathbf{n}^2}} \biggl(10 l_{0,1} \left(3 \left(g_A^2+1\right) m_N^3-g_A^2 m_N M_{\pi}^2\right)\\
&\quad+g_A M_{\pi}^2 \Bigl(5 l_{1,1} \left(3 m_N^2-M_{\pi}^2\right)-10 m_N l_{0,2}\Bigr)\biggr)
\end{split}\\
\begin{split}
\beta_{u-d}^{\text{red}}&=\frac{g_A M_{\pi}^2}{60 \pi ^2 F_{\pi}^2 m_N^3} \Bigl(10 g_A m_N l_{0,1} \left(3 m_N^2-2 M_{\pi}^2\right)+10 l_{0,2} \left(4 m_N^3-m_N M_{\pi}^2\right)\\
&\quad+5 l_{1,1} \left(4m_N^2M_{\pi}^2-M_{\pi}^4\right)\Bigr) 
\end{split}\\
\gamma_{u-d}^{\text{red}}&=-\frac{g_A^2 L M_{\pi}^2 u l_{0,1} \left(m_N^4 (12-9 u)-4 m_N^2 M_{\pi}^2+M_{\pi}^4\right) \mathbf{n}^2}{12 \pi ^2 F_{\pi}^2 m_N^2   \sqrt{\mathbf{n}^2f(u,m_N^2)}}
\end{align}
We show the result of these estimates in fig.~\ref{fig:estimate_1} for several different pion masses. In fig.~\ref{fig:estimate_2} we show the contribution to the finite volume corrections per diagram at $M_{\pi}=250\,\text{MeV}$.\\
We estimate the corrections for reasonable regions of $M_{\pi}L$ to be small and negative, i.e. the lattice data should be corrected towards smaller values of $\langle x\rangle$. For $M_{\pi}L\approx2$ we find a stronger dependence on the input parameters, which leads us to conclude that these formulae are not yet applicable for such small values of $M_{\pi}L$ and indeed, this approach is only valid for $M_{\pi}L\gg1$.\\A thorough analysis of the finite volume corrections to $\langle x\rangle$ will require a complete re-analysis of the lattice data that is available so far. Overall, we estimate this correction to be too small to solve the problems one encounters in the extrapolation of $\langle x\rangle$ towards the physical point \cite{Bratt:2010jn,Bali:2012av}. In particular, our results are in contradiction to the large finite volume effects claimed in \cite{Detmold:2003rq}.
\section{Conclusion\label{sec:conc}}
In this paper, we have calculated the full one-loop finite volume corrections to the quantities $\langle x\rangle_{u\pm d}$ within the framework of covariant BChPT following the methods presented in \cite{Gasser:1986vb,Gasser:1987ah,Gasser:1987zq,Hasenfratz:1989pk,Wein:2011ix}. We then proceeded to give estimates for the vector quantity and we found the finite volume corrections to be small. Presently, an analysis of new lattice data with quark masses down to close to physical values and including these finite volume corrections is in preparation \cite{Bali:2014}.
\appendix
\section{Infrared integrals\label{app:integrals}}
We define our fundamental integrals as proposed in \cite{Becher:1999he} in the following way:
\begin{align}
I_{k,l}(p^2,m^2,M^2)&=\frac{1}{i}\int\frac{d^Dq}{(2\pi)^D}\frac{1}{(M^2-q^2)^k(m^2-(p-q)^2)^l}.
\end{align}
Following the reasoning and definitions given in that paper, one arrives at the following lowest basic integrals:
\begin{align}
I_{1,0}(p^2,m^2,M^2)&=2M^2\lambda+\frac{M^2}{(4\pi)^2}\log\frac{M^2}{\mu^2},\\
I_{0,1}(p^2,m^2,M^2)&=0,\\
\begin{split}
I_{1,1}(p^2,m^2,M^2)&=-\frac{M^2-m^2+p^2}{p^2}\lambda+\frac{1}{(4\pi)^2}\Biggl[\frac{M^2-m^2+p^2}{2p^2}\left(1-\log\frac{M^2}{\mu^2}\right)\\
&\quad-2\sqrt{\frac{M^2}{p^2}-\frac{(M^2-m^2+p^2)^2}{4p^4}}\arccos\frac{-M^2+m^2-p^2}{\sqrt{4M^2p^2}}\Biggr].
\end{split}
\end{align}
Here, we have used the abbreviation $\lambda$ as defined in \cite{Becher:1999he} and $\mu$ is the dimensional regularization scale. Note that higher versions of these integrals can be obtained by taking derivatives with respect to the masses $M^2$ or $m^2$.
\section{Tensor decomposition\label{app:tensor}}
In this chapter we show how we carried out the tensor decomposition for the tensorial infrared integrals. We start off by defining the tensorial integrals:
\begin{align}
I_{k,l}^{\mu\nu \cdots}  \equiv -i \int \frac{d^dq}{(2\pi)^d} \frac{q^\mu q^\nu \cdots}{(M^2-q^2-i \epsilon)^k(m^2-(p-q)^2-i \epsilon)^l}.
\end{align}
Employing Lorentz decomposition, we can write
\begin{subequations}
 \begin{align}
I_{k,l}^{\mu} & \equiv p^\mu I_{k,l}^{(1)} \ , \\
I_{k,l}^{\mu \nu} & \equiv p^2 g^{\mu\nu} I_{k,l}^{(2)} + p^\mu p^\nu I_{k,l}^{(3)} \ , \\
I_{k,l}^{\mu \nu \rho} & \equiv p^2 (g^{\mu\nu} p^\rho + g^{\nu\rho} p^\mu + g^{\rho\mu} p^\nu ) I_{k,l}^{(4)} + p^\mu p^\nu p^\rho I_{k,l}^{(5)} \ , \\
\begin{split}
I_{k,l}^{\mu \nu \rho \sigma} & \equiv p^4 (g^{\mu\nu} g^{\rho\sigma} + g^{\mu\rho} g^{\sigma \nu} + g^{\mu\sigma} g^{\nu\rho} ) I_{k,l}^{(6)}  \\
& \quad + p^2 (p^\mu p^\nu g^{\rho\sigma}+p^\mu p^\rho g^{\sigma\nu}+p^\mu p^\sigma g^{\nu\rho}+p^\rho p^\sigma g^{\mu\nu}+p^\sigma p^\mu g^{\nu\rho}+p^\nu p^\rho g^{\mu\sigma}) I_{k,l}^{(7)}\\
& \quad+ p^\mu p^\nu p^\rho p^\sigma I_{k,l}^{(8)} \ .
\end{split}
 \end{align}
\end{subequations}
In order to write down the explicit expressions for these $I^{(i)}_{k,l}$, let us first introduce two abbreviations:
\begin{align}
T_{k,l}^{(i)} \equiv -I_{k-1,l}^{(i)}+M^2 I_{k,l}^{(i)},\qquad U_{k,l}^{(i)} \equiv  I_{k,l-1}^{(i)}-I_{k-1,l}^{(i)}+(M^2-m^2+p^2)I_{k,l}^{(i)},
\end{align}
since these frequently appear in the explicit forms of these $I^{(i)}_{k,l}$. They take the form (using that $I_{0,0}=0$)
\begin{subequations}
\begin{align}
I_{k,l}^{(1)} & = \frac{1}{2 p^2} U_{k,l}^{(0)} \ , \\
I_{k,0}^{(1)} & = 0 \ , \\
I_{0,l}^{(1)} & = I_{0,l} \ ,
\end{align}
\end{subequations}
\begin{subequations}
 \begin{align}
I_{k,l}^{(2)} & = \frac{1}{(d-1) p^2} \left( T_{k,l}^{(0)} - \frac{1}{2} U_{k,l}^{(1)}\right), \\
I_{k,0}^{(2)} & = \frac{1}{d p^2} \bigl( -I_{k-1,0}+M^2 I_{k,0} \bigr), \\
I_{0,l}^{(2)} & = \frac{1}{d p^2} \bigl( -I_{0,l-1}+m^2 I_{0,l-1} \bigr),
\end{align}
\end{subequations}
\begin{subequations}
 \begin{align}
I_{k,l}^{(3)} & = \frac{-1}{(d-1) p^2} \left( T_{k,l}^{(0)} - \frac{d}{2} U_{k,l}^{(1)}\right), \\
I_{k,0}^{(3)} & = 0, \\
I_{0,l}^{(3)} & = I_{0,l},
\end{align}
\end{subequations}
\begin{subequations}
 \begin{align}
I_{k,l}^{(4)} & = \frac{1}{(d-1) p^2} \biggl( T_{k,l}^{(1)}  - \frac{1}{2} \Bigl( U_{k,l}^{(2)}+U_{k,l}^{(3)} \Bigr) \biggr), \\
I_{k,0}^{(4)} & = 0, \\
I_{0,l}^{(4)} & =\frac{1}{d p^2} \bigl( -I_{0,l-1}+m^2 I_{0,l-1} \bigr),
\end{align}
\end{subequations}
\begin{subequations}
 \begin{align}
I_{k,l}^{(5)} & = \frac{-1}{(d-1) p^2} \biggl( 3 T_{k,l}^{(1)} - \frac{d+2}{2} \Bigl( U_{k,l}^{(2)}+U_{k,l}^{(3)} \Bigr) \biggr), \\
I_{k,0}^{(5)} & = 0, \\
I_{0,l}^{(5)} & = I_{0,l},
\end{align}
\end{subequations}
\begin{align}
I_{k,l}^{(6)} & = \frac{1}{(d+1) p^2} \left( T_{k,l}^{(2)} - \frac{1}{2} U_{k,l}^{(4)}\right),
\end{align}
\begin{align}
I_{k,l}^{(7)} & = \frac{-1}{(d+1) p^2} \left( T_{k,l}^{(2)} - \frac{d+2}{2} U_{k,l}^{(4)}\right),
\end{align}
\begin{align}
\begin{split}
 I_{k,l}^{(8)} & = \frac{1}{p^2} T_{k,l}^{(3)} - (d+4) I_{k,l}^{(7)} = \frac{-1}{(d+1) p^2} \left( 3 T_{k,l}^{(3)} - \frac{d+4}{2} U_{k,l}^{(5)}\right).
\end{split}
\end{align}
\acknowledgments{This work was supported by the Deutsche Forschungsgemeinschaft SFB/Transregio 55.}
\bibliographystyle{apsrev}

\end{document}